\documentclass[journal=aamick,manuscript=article]{achemso}
\usepackage[version=4]{mhchem} % Formula subscripts using \ce{}
\usepackage{tabularray}
\usepackage{amsmath}
\usepackage{bm}
\usepackage{epsfig}
\usepackage{epstopdf}
\usepackage{multirow}
\usepackage{graphicx}
\usepackage{subcaption}
\usepackage{siunitx}
\author{Vuong Van Thanh}
\email{thanh.vuongvan@hust.edu.vn}
\affiliation{School of Mechanical Engineering, Hanoi University of Science and Technology, Hanoi 100000, Viet Nam}
\author{Nguyen Minh Quan}
\affiliation{School of Mechanical Engineering, Hanoi University of Science and Technology, Hanoi 100000, Viet Nam}
\author{Nguyen Tuan Hung}
\email{nguyenth@ntu.edu.tw}
\affiliation{Department of Materials Science and Engineering, National Taiwan University, Taipei 106319, Taiwan}

\title[title]{Tunable Rashba Splitting in Janus InXPbP (X = S, Se, Te) Monolayers for Enhanced Photocatalytic Water Splitting}
\abbreviations{}

\begin{document}

\begin{abstract}
% Câu đầu tiên giới thiệu về Rashba (SOC)
Janus two-dimensional (2D) materials exhibiting Rashba spin splitting have recently attracted considerable attention owing to their potential applications in spintronic devices and photocatalytic water splitting. In this work, we investigate, using first-principles calculations, the structural, mechanical, electronic, optical, and photocatalytic properties of Janus InXPbP (X = S, Se, Te) monolayers that exhibit significant Rashba effects. Our results demonstrate that all three monolayers are energetically, dynamically, and mechanically stable, as evidenced by cohesive energy calculations, phonon dispersion analysis, and elastic constants. By varying the chalcogen atom ($X=\mathrm{S,Se,Te}$), the Rashba effect in InXPbP can be effectively tuned. Rashba parameters of 0.16 and 0.20~eV\AA{} are obtained near the conduction-band minimum (CBM) for InSPbP and InSePbP, respectively, whereas InTePbP exhibits giant Rashba spin splitting near both the CBM and valence-band maximum (VBM), with corresponding Rashba parameters of 0.90 and 0.87~eV\AA{}. Furthermore, the Janus InXPbP monolayers exhibit suitable band gaps of 1.21, 1.27, and 0.76 eV for InSPbP, InSePbP, and InTePbP, respectively, which are favorable for photocatalytic applications. All three monolayers possess suitable band-edge alignments for overall water splitting, yielding solar-to-hydrogen (STH) conversion efficiencies of 21.67\%, 26.03\%, and 29.83\% for InSPbP, InSePbP, and InTePbP, respectively. Our findings not only enrich the family of Janus materials but also suggest that the Janus InXPbP monolayers are promising candidates for spintronic devices and high-performance photocatalytic water-splitting applications.
\end{abstract}

%\begin{tocentry}

%\begin{center}
%\includegraphics[width=0.70\textwidth]{TOC.pdf}
%\end{center}

%\end{tocentry}

%%%MAIN TEXT%%%%
\section{Introduction}
The rapid growth in global energy demand, together with increasingly severe environmental pollution from fossil fuel consumption, has strongly stimulated research on clean and sustainable energy technologies~\cite{chodvadiya2021enhancement,thanh2023janus}. Owing to its high energy density, carbon-free operation, and broad potential for future applications, hydrogen is regarded as one of the most promising energy carriers for sustainable energy systems~\cite{roy2024review}. Among current hydrogen production methods, photocatalytic water splitting has been widely studied for its ability to directly utilize solar energy to generate clean and renewable chemical fuels~\cite{fujishima1972electrochemical,trang2024manipulating}.

Recently, two-dimensional (2D) materials have been extensively investigated for photocatalytic applications owing to their large surface-to-volume ratios and remarkable electronic, optical, and mechanical characteristics~\cite{fu2018intrinsic, bouziani2023computational}. One of the fundamental criteria for photocatalytic water splitting is that the photocatalyst must possess a band gap larger than the energy difference between the water redox potentials (1.23 eV)~\cite{luo2021pd4s3se3,din2019rashba}. However, Janus materials can potentially drive water splitting even with band gaps below 1.23~eV owing to their intrinsic polarization and built-in electric fields, which promote efficient charge separation and facilitate the redox reactions~\cite{luo2021pd4s3se3,thanh2023janus,bao2025rational}. For instance, several Janus monolayers have demonstrated promising photocatalytic performance, including Janus WSSe with an STH efficiency of 11.7\%~\cite{ju2020janus}, as well as $\gamma$-Ge$_2$SSe (28.8\%)~\cite{thanh2023janus}, B$_2$P$_6$ (28.2\%)~\cite{sun2020b2p6}, and Pd$_4$S$_3$Te$_3$ (38.6\%)~\cite{luo2021pd4s3se3}. Nevertheless, further improvement in photocatalytic efficiency is still highly desirable, particularly through mechanisms that can effectively suppress electron--hole recombination and prolong carrier lifetime~\cite{din2019rashba}. 

Recently, the Rashba effect has emerged as a promising strategy for enhancing the photocatalytic performance of Janus materials~\cite{zhang2016spin,din2019rashba}. The Rashba spin splitting induced by strong spin--orbit coupling (SOC) and structural asymmetry can suppress electron--hole recombination by generating momentum and spin splitting near the band edges, thereby prolonging carrier lifetimes and improving charge separation efficiency. However, achieving both pronounced Rashba spin splitting and high photocatalytic water-splitting efficiency simultaneously remains challenging, as many previously reported Janus materials with sizable Rashba effects still exhibit limited photocatalytic performance~\cite{yang2025self}. For instance, Janus MoSSe and WSSe monolayers were predicted to achieve photocatalytic water-splitting efficiencies of 20.39\%~\cite{yang2025self} and 11.7\%, respectively. Therefore, developing multifunctional Janus 2D materials that simultaneously possess pronounced Rashba spin splitting and efficient photocatalytic activity remains a significant challenge. Based on first-principles calculations, Chaoui \textit{et al.}~\cite{chaoui2025novel} proposed Janus SGa--PbP and SeGa--PbP monolayers with promising potential for photocatalytic water splitting. They reported maximum STH conversion efficiencies of 40.69\% for SGa--PbP and 31.75\% for SeGa--PbP. Since In and Ga belong to the same group-III family and possess similar valence electronic configurations, replacing Ga with the heavier In atom is expected to preserve the favorable electronic characteristics while further enhancing the SOC strength. This motivates the exploration of Janus InSPbP, InSePbP and InTePbP systems as potential multifunctional materials for photocatalytic and spintronic applications. In particular, Janus InSPbP, InSePbP, and InTePbP monolayers are promising candidates for photocatalytic water-splitting applications because of their intrinsic structural asymmetry together with the incorporation of heavy elements such as Pb and Te, which are expected to induce strong SOC and sizable Rashba spin splitting. Moreover, replacing the chalcogen atom from S to Se and Te offers an effective strategy for tuning the electronic structure, band-edge alignment, and optical absorption properties associated with photocatalytic water splitting. These features make Janus InSPbP, InSePbP, and InTePbP monolayers attractive platforms for investigating the interplay between Rashba effects and photocatalytic performance.

Here, we systematically investigate the structural, mechanical, electronic, Rashba spin splitting, and photocatalytic properties of Janus InXPbP ($X=\mathrm{S, Se, Te}$) monolayers via density functional theory (DFT) calculations. Our results reveal that these monolayers possess dynamical and mechanical stability, pronounced Rashba spin splitting, strong visible-light absorption, and favorable band-edge alignments for photocatalytic water splitting. These results suggest that Janus InXPbP monolayers are promising candidates for coupled spintronic and energy-conversion applications.

\section{Methodology}
In this study, we employ DFT calculations as implemented in the Quantum ESPRESSO package~\cite{giannozzi2009,nguyen2022QE}. The Perdew-Burke-Ernzerhof (PBE) functional is selected for the exchange--correlation potential~\cite{perdew1996generalized} within the generalized gradient approximation (GGA). Optimized norm-conserving Vanderbilt pseudopotentials~\cite{hamann2013optimized} are used to describe the electron--ion interactions. To obtain more reliable electronic band structures and band gap values, we adopt the Heyd--Scuseria--Ernzerhof (HSE) hybrid functional~\cite{heyd2003hybrid}, and we interpolate the HSE band structures using the Wannier90 package~\cite{mostofi2014updated}. The SOC effects are accounted for in our calculations. According to the convergence tests, the plane-wave basis set is expanded with an energy cutoff of 80~Ry, and a $12 \times 12 \times 1$ $\bm{k}$-point mesh is employed for Brillouin-zone sampling. A vacuum spacing of 24~\AA{} is employed perpendicular to the monolayer plane to avoid interactions between periodic cells due to periodic boundary conditions in the DFT method. Density-functional perturbation theory (DFPT) is employed to compute the phonon dispersions~\cite{baroni2001phonons}. Geometry optimization is performed using the Broyden--Fletcher--Goldfarb--Shanno algorithm~\cite{nguyen2022QE}, with convergence criteria of $1 \times 10^{-5}$~Ry/a.u. for forces and $5 \times 10^{-2}$~GPa for stresses. A dense $48 \times 48 \times 1$ Monkhorst--Pack $k$-point mesh and an interband broadening parameter of 0.25~eV (\texttt{intersmear}=0.25) were employed in the optical calculations to obtain well-converged dielectric spectra.

\section{Results and discussion}

\subsection{Geometric Structure and Stability.}

\begin{figure*}[t]
  \centering \includegraphics[clip,width=12cm]{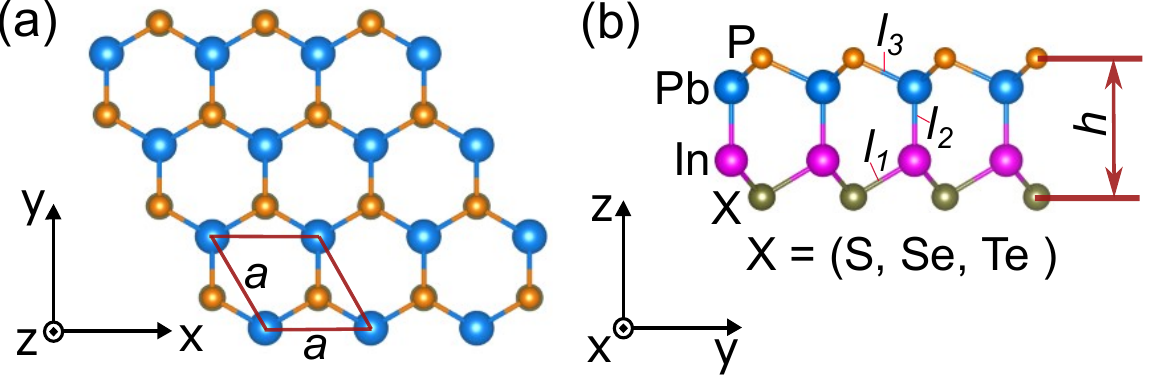}
  \caption{(a) Top view and (b) side view of the Janus InXPbP (X=S, Se, Te) monolayers.}
  \label{fig:model}
\end{figure*}

In Figs.~\ref{fig:model}(a) and (b), we show the top and side views of the atomic structures of Janus InXPbP (X= S, Se, Te) monolayers. The optimized lattice constants for InSPbP, InSePbP, and InTePbP are 4.04, 4.10, and 4.23~\AA{}, respectively, as listed in Table~1. These values are generally larger than those reported for related Janus and group-III monochalcogenide systems, such as InSe monolayers (3.90~\AA{})~\cite{hung2017two}, GaS--PbP (3.87~\AA{}), and GeSe--PbP (3.95~\AA{})~\cite{chaoui2025novel}. The lattice constant increases monotonically from InSPbP to InTePbP due to the progressively larger size of the chalcogen species from S to Te, which results in an expansion of the in-plane lattice. Consistently, the X (S, Se, Te)--In ($l_{1}$), In--Pb ($l_{2}$), and Pb--P ($l_{3}$) bond lengths increase from InSPbP to InTePbP due to the progressively larger atomic size of the chalcogen atoms, which reduces orbital overlap. In particular, $l_{1}$ exhibits the most pronounced variation, whereas $l_{2}$ remains nearly unchanged, indicating the anisotropic response of different bonding environments within the Janus structure. Moreover, the buckling height $h$ increases from 5.284~\AA{} for InSPbP to 5.567~\AA{} for InTePbP, indicating enhanced out-of-plane structural distortion with increasing chalcogen atomic size. A similar trend is observed for the effective layer thickness $d_{0}$, which increases from 9.074~\AA{} to 9.457~\AA{}, further reflecting the vertical expansion of the monolayers. Here, $d_{0}$ is defined as $d_{0}=h+r_{X}+r_{P}$, where $h$ denotes the buckling height, while $r_{X}$ and $r_{P}$ are the van der Waals radii of the X (S, Se, Te) and P atoms~\cite{alvarez2013cartography}, respectively.

\begin{table*}[t]
\centering
\caption{Lattice constant $a$ (\AA), bond lengths $l_{1}$, $l_{2}$, and $l_{3}$ (\AA), buckling height $h$ (\AA), effective layer thickness $d_0$ (\AA), elastic constants $C_{ij}$ (N/m), and conhesive energy $E_{\text{coh}}$ (eV/atom)}

%\vspace{0.5cm}  % Thêm khoảng cách 0.5cm
%\small
%\footnotesize  % cỡ chữ trong bảng
\renewcommand{\arraystretch}{1.5}
\begin{tabular}{c c c c c c c c}
\hline\hline
Materials & $a$  & $l_{1}$  & $l_{2}$  &  $l_{3}$ & $h$ & $d_{0}$ & $E_{coh}$ \\\hline
InSPbP & 4.04 & 2.58 & 2.94 & 2.64 & 5.28 & 9.07 & -3.42    \\  
InSePbP & 4.10 & 2.68 & 2.94 & 2.66 & 5.42 & 9.14 & -3.23 \\ 
InTePbP & 4.23 & 2.85 & 2.94 & 2.70 & 5.57 & 9.46 & -3.19 \\ 
\hline\hline
\end{tabular}
\label{tab:table1}
\end{table*}    

To examine their potential experimental realizability, the cohesive energies, $E_{\text{coh}}$ in the unit of eV per atom, of Janus InXPbP (X = S, Se, Te) monolayers are calculated using the following equation~\cite{}:
\begin{equation}
\label{eq:coh}
E_{\text{coh}} = \frac{E_{\text{tot}} - \left( E_{\text{In}} + E_{\text{X}} + E_{\text{Pb}} + E_{\text{P}} \right)}{4},
\end{equation}
where $E_{\text{tot}}$ is the total energy of the Janus InXPbP monolayer unit cell, and $E_{\text{In}}$, $E_{\text{X}}$ (X = S, Se, Te), $E_{\text{Pb}}$, and $E_{\text{P}}$ correspond to the energies of isolated In, chalcogen, Pb, and P atoms, respectively. The calculated cohesive energies of the Janus InSPbP, InSePbP, and InTePbP monolayers are $-3.42$, $-3.23$, and $-3.19$ eV/atom, respectively. The reduction in cohesive energy from InSPbP to InTePbP indicates a slight weakening of interatomic interactions. The relatively large negative cohesive energies confirm strong interatomic bonding within the system, ensuring the structural stability of the Janus InXPbP monolayers and suggesting their feasibility for experimental realization.

\begin{figure*}[t]
\centering \includegraphics[clip,width=17cm]{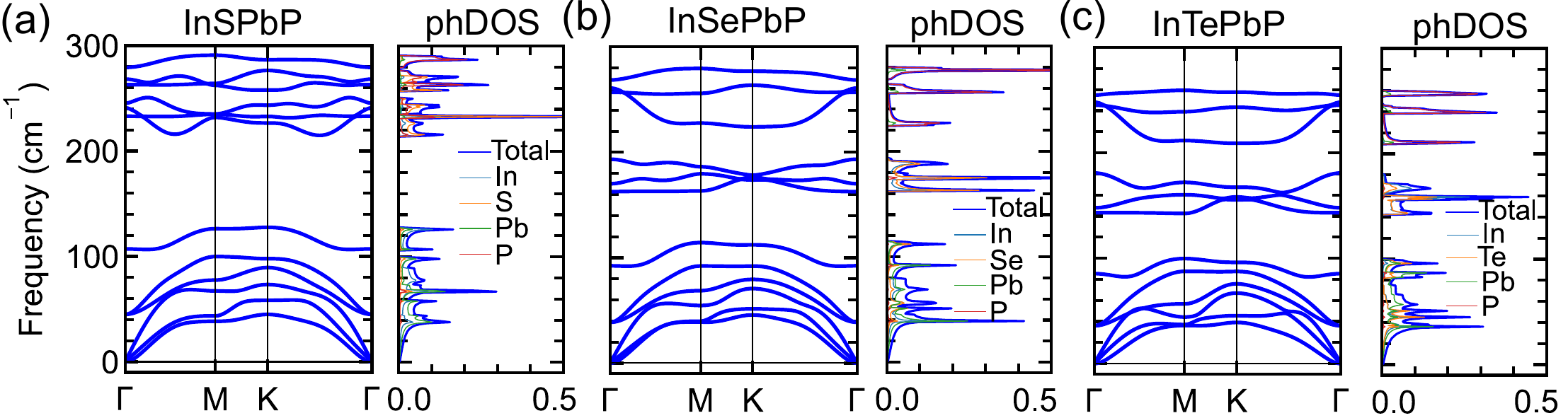}
\caption{Phonon dispersion and phonon density of states (phDOS) for Janus (a) InSPbP, (b) InSePbP, and (c) InTePbP monolayers.}
\label{fig:phonon}
\end{figure*}

Next, we calculated the phonon spectra to evaluate the dynamical stability of the studied materials. Our results indicate that no imaginary phonon modes appear throughout the entire Brillouin zone for all three structures, as shown in Fig.~\ref{fig:phonon}, thereby confirming their dynamical stability. In addition, we examine the mechanical stability of the investigated monolayers by calculating the elastic constants $C_{ij}$ ($C_{11}$, $C_{12}$, and $C_{66}$). The calculated $C_{ij}$ values of the Janus InXPbP monolayers are listed in Table~\ref{tab:table2}. The calculated values fulfill the Born mechanical stability conditions ($C_{11} > |C_{12}| > 0$ and $C_{66} > 0$)~\cite{mouhat2014necessary}, indicating the mechanical stability of all three investigated structures. Thus, the Janus InXPbP monolayers are mechanically stable.

\subsection{Mechanical Properties}

After confirming the dynamical and mechanical stability of Janus InXPbP (X = S, Se, Te) monolayers, we further examine their mechanical behavior. Understanding these properties is essential for assessing the potential of these materials in flexible-device applications. The Young's modulus $Y$ and Poisson's ratio $\upsilon$ are determined from the following expressions~\cite{thanh2021charge,Thanh2023}:
\begin{equation}
  \label{eq:young}
    Y_{xx}=\frac{C_{11}C_{22}-C_{12}^{2}}{C_{22}};~Y_{yy}=\frac{C_{11}C_{22}-C_{12}^{2}}{C_{11}}
\end{equation}
and 
\begin{equation}
  \label{eq:poisson}
       \upsilon_{xy}=\frac{C_{12}}{C_{22}};~\upsilon_{yx}=\frac{C_{12}}{C_{11}}.
\end{equation}

\begin{table}[t]
\caption{Elastic constants $C_{ij}$ (N/m), Young’s modulus $Y_{xx(yy)}$ (N/m), Poisson’s ratio $\nu_{xy(yx)}$, and ideal strength $\sigma$ (N/m) with corresponding strain $\epsilon$.}
\centering
%\small
\renewcommand{\arraystretch}{1.5}
\begin{tabular}{lcccccccc}
\hline
Material & $C_{11(22)}$ & $C_{12}$ & $C_{66}$ & $Y_{xx(yy)}$ & $\nu_{xy(yx)}$ & $\sigma_{xx}$ ($\epsilon_{xx}$) & $\sigma_{yy}$ ($\epsilon_{yy}$) & $\sigma_{bia}$ ($\epsilon_{bia}$) \\
\hline
InSPbP  & 58.6 & 14.8 & 21.6 & 53.0 & 0.26 & 5.32 (0.20) & 6.64 (0.24) & 6.51 (0.22) \\
InSePbP & 55.3 & 15.0 & 20.0 & 51.2 & 0.27 & 4.97 (0.20) & 6.36 (0.26) & 5.87 (0.22) \\
InTePbP & 53.1 & 11.9 & 19.3 & 50.4 & 0.22 & 4.17 (0.18) & 5.41 (0.26) & 5.16 (0.20) \\
\hline
\end{tabular}
\label{tab:table2}
\end{table}
% viết lại đoạn này ...

The calculated Young's modulus values, $Y_{xx(yy)}$, are 53.0, 51.2, and 50.4 N/m for InSPbP, InSePbP, and InTePbP, respectively. The corresponding Poisson's ratios, $\nu_{xy(yx)}$, are 0.26, 0.27, and 0.22. Compared with graphene (336.3 N/m)~\cite{van2019charge}, striped borophene (376.6 N/m)~\cite{thanh2021charge}, monolayer Janus MoSSe (123.4 N/m)~\cite{van2020first}, Janus SGa--PbP (59.04 N/m), SeGa--PbP (56.99 N/m)~\cite{chaoui2025novel}, and AlGaTe$_2$ (56.23 N/m)~\cite{guo2025janus}, the lower Young's modulus values of the present systems suggest enhanced mechanical flexibility. The calculated elastic parameters are summarized in Table~\ref{tab:table2}.

\begin{figure*}[t] 
  \centering \includegraphics[clip,width=16cm]{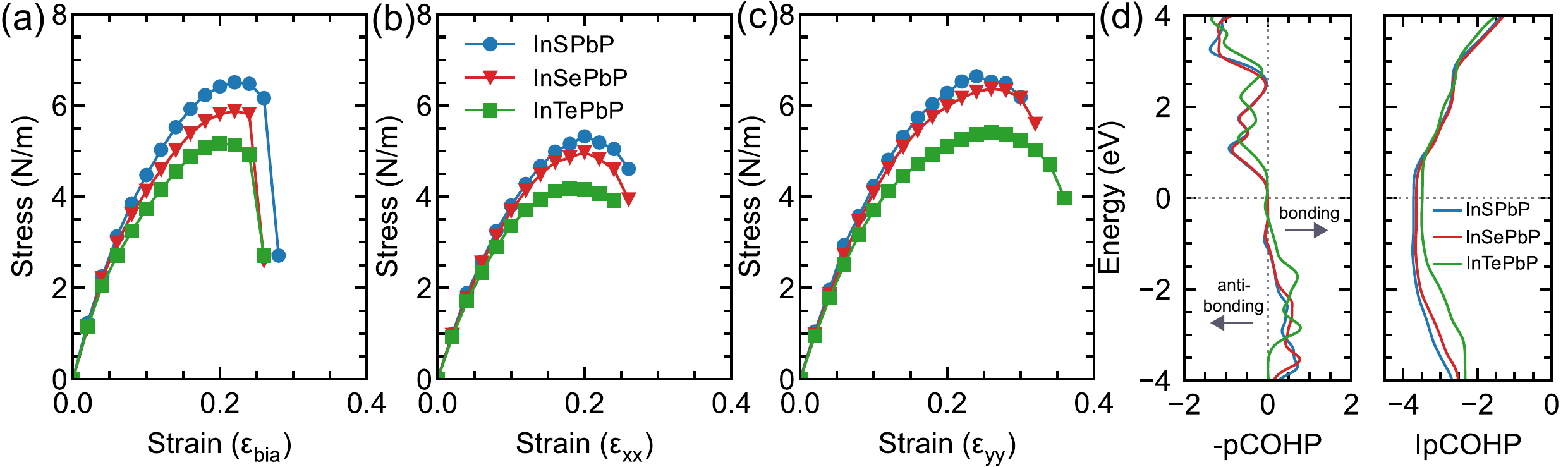}
\caption{Stress--strain curves of Janus InXPbP monolayers under (a) biaxial strain and uniaxial strain applied along the (b) $x$- and (c) $y$-directions. (d) Projected crystal orbital Hamilton population (pCOHP) and integrated pCOHP (IpCOHP).}
  \label{fig:mechanical}
\end{figure*}

In Figs.~\ref{fig:mechanical}(a--c), we show the stress--strain curves of Janus InXPbP ($X=\mathrm{S,Se,Te}$) monolayers under (a) biaxial strain, (b) uniaxial tensile strain along the $x$ direction, and (c) uniaxial tensile strain along the $y$ direction. We find that the mechanical strength decreases from InSPbP to InTePbP under both uniaxial and biaxial loading conditions. This behavior can be attributed to the decreasing electronegativity of the chalcogen atoms, where S (2.58 on the Pauling scale) is more electronegative than Se (2.55) and Te (2.10). Higher electronegativity enhances bonding interactions~\cite{hung2025strain}, leading to shorter bond lengths and a more compact lattice structure. As a result, the stronger bonding in InSPbP contributes to its higher stiffness and mechanical strength compared to the other systems. This trend can be further understood from the bonding perspective by analyzing the IpCOHP. As shown in Fig.~\ref{fig:mechanical}(d), the IpCOHP values of InSPbP are more negative than those of InSePbP and InTePbP, indicating stronger interatomic bonding due to a higher occupation of bonding states. Compared with representative 2D materials, the ideal strength of InSPbP along the $y-$ direction (6.64 N/m) is lower than those reported for representative 2D materials, including MoS$_2$ (15.1 $\pm$ 3 N/m)~\cite{bertolazzi2011stretching}, borophene (12.1 N/m)~\cite{thanh2021charge}, and Janus B$_2$P$_6$ (18.6 N/m)~\cite{van2024janus}.

\subsection{Electronic Properties}

\begin{figure*}[!t]   
  \centering \includegraphics[clip,width=13.5cm]{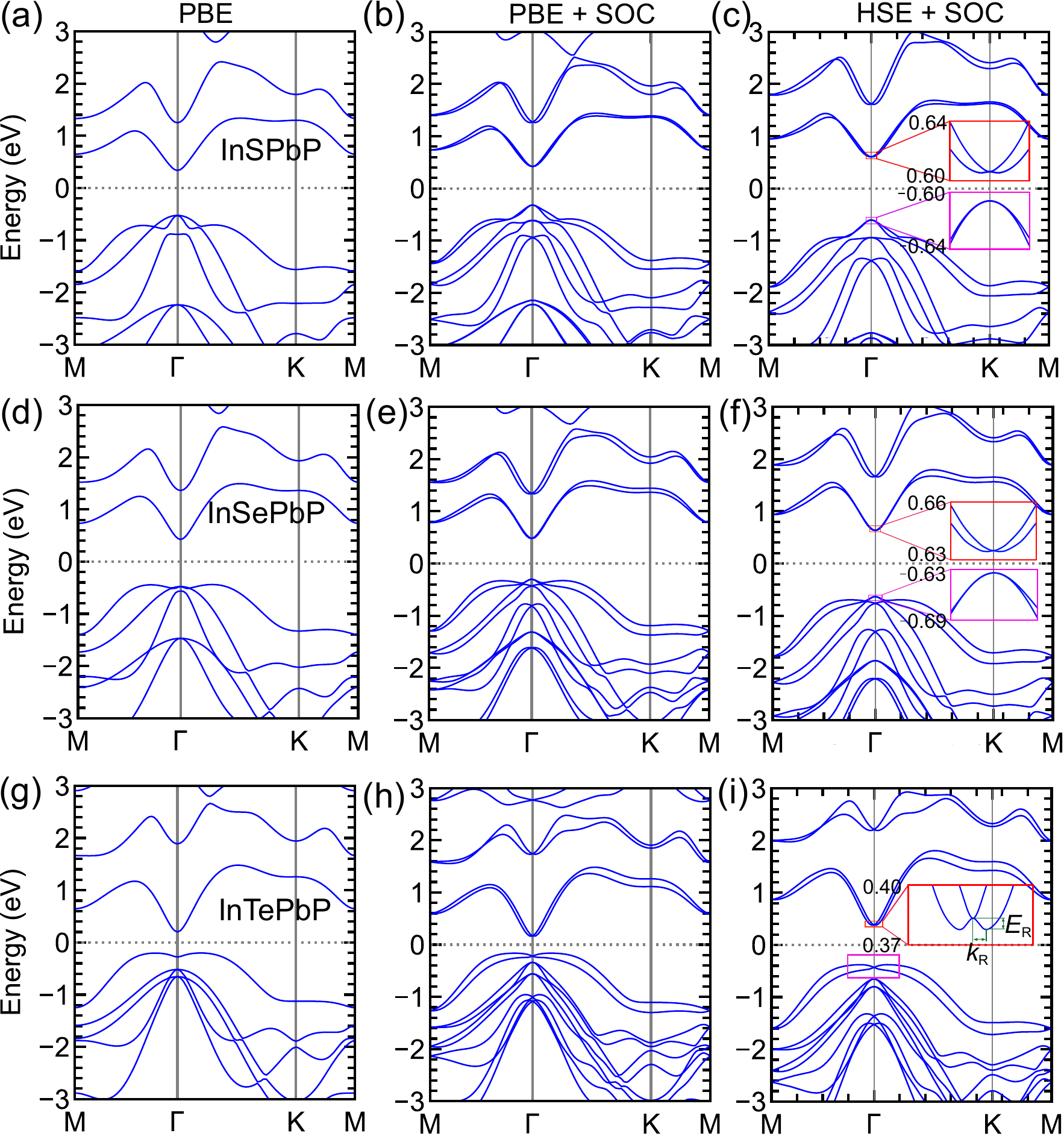}
 \caption{Band structures of Janus InSPbP, InSePbP, and InTePbP monolayers are shown in panels (a--c), (d--f), and (g--i), respectively, calculated using the PBE, PBE+SOC, and HSE + SOC methods. The red rectangle in panel (i) highlights the Rashba-type spin splitting near the band extrema.}
\label{fig:bands}
\end{figure*}

% bỏ PDOS 
%\begin{figure*}[!t]   
%  \centering 
%  \includegraphics[clip,width=13.0cm]{pdos-all.pdf}
%  \caption{PDOS of Janus (a) InSPbP, (b) InSePbP, and (c) InTePbP monolayers.}
%\label{fig:pdos-all}
%\end{figure*}

%%% Thêm trích dẫn tài liệu band gap của một số vật liệu đưa vào tham khảo. ....

In Fig.~\ref{fig:bands}, we show the electronic band structures of Janus InSPbP, InSePbP, and InTePbP obtained using the PBE, PBE+SOC, and HSE+SOC approaches, in which Fig.~\ref{fig:bands}(a--c), (d--f), and (g--i) correspond to InSPbP, InSePbP, and InTePbP, respectively. The calculated results indicate that InSPbP exhibits a direct band gap at the PBE level but undergoes a direct-to-indirect band-gap transition upon inclusion of SOC. Meanwhile, InSePbP and InTePbP remain indirect-gap semiconductors both with and without SOC. These band-gap characteristics are consistently reproduced by the HSE+SOC calculations. The corresponding band-gap values obtained from the PBE, PBE+SOC, and HSE+SOC calculations are summarized in Table~\ref{tab:table3}. In particular, the HSE+SOC calculations, which generally provide a more reliable description of electronic band gaps, predict values of 1.21, 1.27, and 0.76 eV for InSPbP, InSePbP, and InTePbP, respectively. These moderate band-gap values are favorable for visible-light absorption and fall within the range commonly reported for 2D semiconductors with promising photocatalytic performance, including Janus $\gamma-$Ge$_{2}$SSe (1.13 eV)~\cite{Thanh2023}, Janus SGa--PbP (0.93 eV) and SeGa--PbP (1.24 eV)~\cite{chaoui2025novel}. Interestingly, the inclusion of SOC gives rise to Rashba spin splitting near the CBM in all three Janus InXPbP monolayers. In particular, InTePbP exhibits additional Rashba splitting near the VBM, as clearly illustrated in Fig.~\ref{fig:bands}(i). These results demonstrate that SOC plays a dual role in these systems by modifying the band-gap characteristics and generating spin-dependent electronic states.
%conduction-band minimum (CBM); valence-band maximum (VBM)
% đoạn trên chỉ rõ điểm CBM, VBM ở vùng nào để thấy được indirect or direct ...
%Thêm giá trị band gap và trích dẫn ở đoạn này ...

\begin{table*}[t]
\centering
\caption{Calculated band gaps $E_g$ (eV) obtained using PBE, PBE+SOC, and HSE + SOC functionals, together with the macroscopic potential difference $\Delta V$ (eV). Rashba coefficients $\alpha_R$ (eV\AA{}) at the CBM and VBM are presented for Janus InXPbP monolayers with $X = \mathrm{S, Se, Te}$.}
%\vspace{0.5cm}  % Thêm khoảng cách 0.5cm
%\small
%\footnotesize  % cỡ chữ trong bảng
\renewcommand{\arraystretch}{1.5}
\begin{tabular}{c c c c c c c c c}
\hline\hline
Materials & $E_{g}^\text{PBE}$ & $E_{g}^\text{PBE + SOC}$ & $E_{g}^\text{HSE + SOC}$ & $\Delta V$ & $\alpha_R^\text{CBM}$ & $\alpha_R^\text{VBM}$ \\\hline
InSPbP & 0.86 & 0.74 & 1.21 & 0.22 & 0.16 & 0.00 \\  
InSePbP & 0.86 & 0.78 & 1.27 & 0.38 & 0.20 & 0.00 \\ 
InTePbP & 0.40 & 0.33 & 0.76 & 0.64 & 0.90 & 0.87\\ \hline\hline
\end{tabular}
\label{tab:table3}
\end{table*}    

To characterize the Rashba spin splitting, we evaluate the Rashba coefficient for InSPbP and InTePbP using $\alpha_R = 2E_R/k_R$~\cite{hu2018intrinsic,wang2017wide,david2026rashba}, where $E_R$ and $k_R$ denote the splitting energy and momentum offset, respectively, as schematically illustrated in Fig.~\ref{fig:bands}(i). Accordingly, the Rashba coefficients $\alpha_R$ are determined to be 0.16 and 0.20~eV\AA\ for the CBM, while $\alpha_R \approx 0$ eV\AA\ for the VBM of InSPbP and InSePbP, respectively. InTePbP exhibits remarkably large Rashba parameters of 0.90 and 0.87~eV\AA\ at the CBM and VBM, respectively. In particular, $\alpha_R$ of InTePbP is considerably larger than those reported for several representative Janus materials, such as MoSSe (0.077 eV$\text{\AA}$), and MoSTe (0.147 eV$\text{\AA}$)~\cite{hu2018intrinsic}, as well as Janus In$_{2}$TeSe monolayer (0.44 eV$\text{\AA}$)~\cite{david2026rashba}, highlighting its strong spin--orbit-driven electronic response.

\begin{figure*}[t]   
  \centering 
  \includegraphics[width=13.5cm]{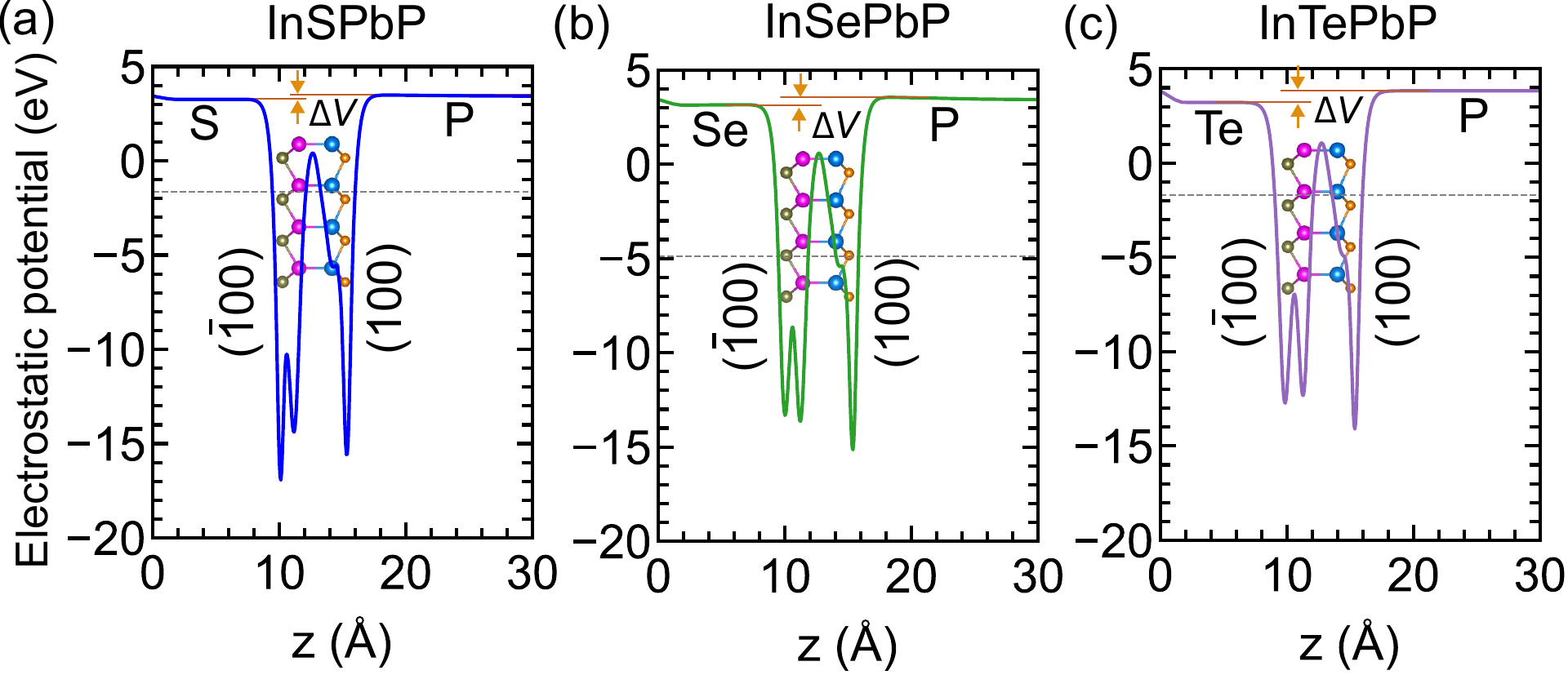}
 \caption{Planar-averaged electrostatic potentials with dipole corrections for Janus (a) InSPbP, (b) InSePbP, and (c) InTePbP monolayers. The dashed line denotes the Fermi level.}
\label{fig:potential-all}
\end{figure*}

To further elucidate the origin of the pronounced Rashba spin splitting, we analyze the planar-averaged electrostatic potential profiles shown in Figs.~\ref{fig:potential-all}(a--c) for Janus InSPbP, InSePbP, and InTePbP monolayers. The calculations include dipole correction to eliminate artificial electric fields arising from periodic boundary conditions by introducing a compensating ramp potential~\cite{nguyen2022QE}. The electrostatic potential differences ($\Delta V = 0.22$, 0.38, and 0.64~eV), which are summarized in Table~\ref{tab:table2}, increase systematically from InSPbP to InTePbP. This trend indicates an increase in the intrinsic out-of-plane electric field and closely tracks the evolution of the Rashba coefficient. In particular, InTePbP exhibits both the largest $\Delta V$ and the strongest $\alpha_R$ compared with InSPbP and InSePbP. Together with the stronger SOC associated with the heavier Te atom, the enhanced internal electric field gives rise to the exceptionally large Rashba parameter observed in InTePbP. Furthermore, the built-in electric field can facilitate the separation of photogenerated electron--hole pairs, which is beneficial for photocatalytic water splitting~\cite{thanh2023janus,bao2025rational}.

\begin{table}[t]
\centering
\caption{Calculated Bader charge ($q$) and associated charge transfer ($\Delta q$) in units of the elementary charge (\(e\)) for InXPbP (X = S, Se, Te) monolayers.}
\renewcommand{\arraystretch}{1.5}
\begin{tabular}{lcccccccc}
\hline\hline
 & \multicolumn{4}{c}{Bader charge ($q$)} & \multicolumn{4}{c}{Charge transfer ($\Delta q$)} \\
\cline{2-5} \cline{6-9}
Material & In & X & Pb & P & In & X & Pb & P \\
\hline
InSPbP  & 12.153 & 6.824 & 13.562 & 5.463 & $-0.847$ & 0.824 & $-0.438$ & 0.463 \\
InSePbP & 12.245 & 6.690 & 13.600 & 5.467 & $-0.755$ & 0.690 & $-0.400$ & 0.467 \\
InTePbP & 12.438 & 6.493 & 13.634 & 5.435 & $-0.562$ & 0.493 & $-0.366$ & 0.435 \\
\hline\hline
\label{tab:table4}
\end{tabular}
\end{table}

From the electronegativity perspective, S (2.58 on the Pauling scale) is more electronegative than Se (2.55) and Te (2.10), resulting in a larger charge transfer from neighboring atoms. As summarized in Table~\ref{tab:table4}, the charge transfer (in units of the elementary charge $e$) to the chalcogen atom decreases from $\Delta q_{\mathrm{S}} = 0.824e$ to $\Delta q_{\mathrm{Se}} = 0.690e$ and further to $\Delta q_{\mathrm{Te}} = 0.493e$. Moreover, the charge-transfer difference between the two outer surfaces, quantified by $|\Delta q_\mathrm{X}-\Delta q_\mathrm{P}|$, decreases from 0.361$e$ in InSPbP to 0.223$e$ in InSePbP and 0.058$e$ in InTePbP, indicating a reduction in the local charge asymmetry between the two sides of the Janus structure. Interestingly, the evolution of the Rashba parameter does not follow the trend of charge transfer obtained from the Bader analysis, highlighting that the magnitude of charge transfer alone is insufficient to describe the built-in electric field in these systems. In contrast, the Rashba coefficient exhibits the same increasing trend as the electrostatic potential difference $\Delta V$. This correlation suggests that the Rashba-related internal electric field is governed not only by the amount of transferred charge but also by its spatial distribution across the monolayer. Furthermore, the stronger SOC associated with the heavier Te atom further amplifies the spin splitting, leading to the exceptionally large Rashba parameter observed in InTePbP.

\subsection{Photocatalytic Water Splitting}

\begin{figure*}[t]
\centering
\includegraphics[width=\linewidth]{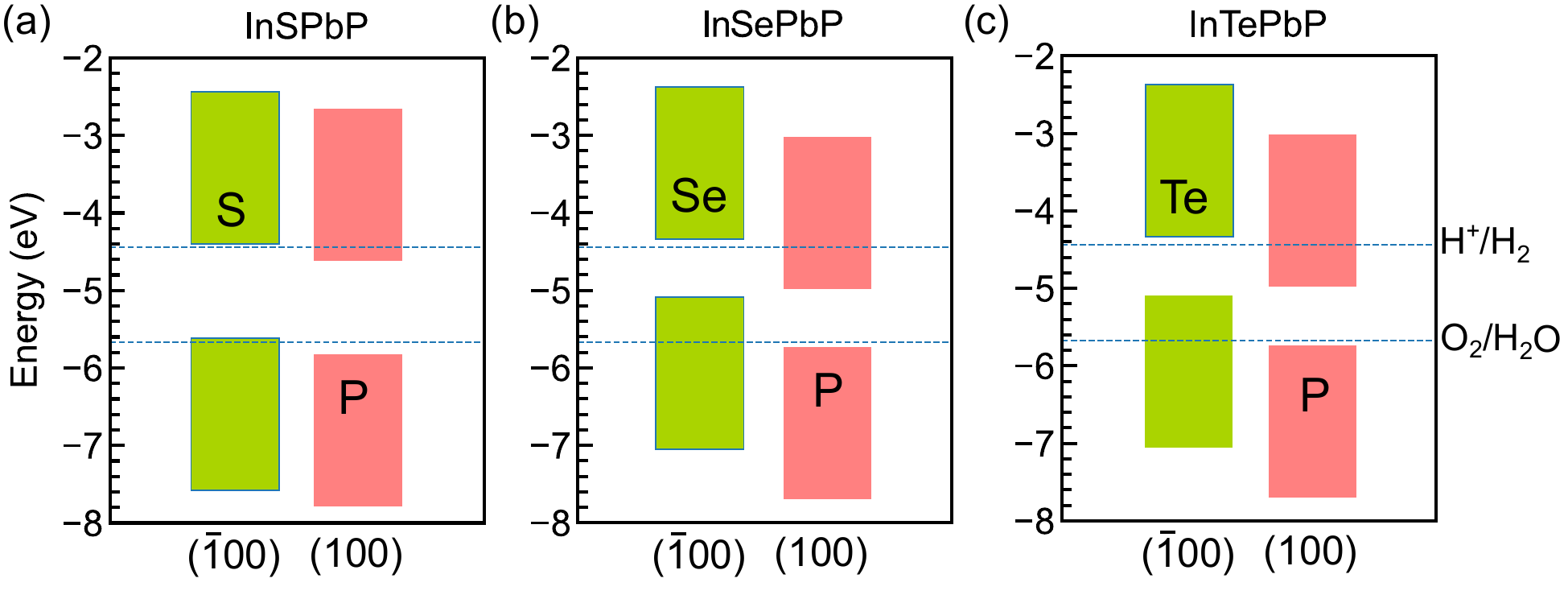}
\caption{Band edge positions of the Janus (a) InSPbP, (b) InSePbP, and (c) InTePbP monolayers with respect to the vacuum level. The redox potentials for hydrogen evolution ($\mathrm{H^+/H_2}$) and oxygen evolution ($\mathrm{O_2/H_2O}$) at pH = 0.}
\label{fig:band-edge}
\end{figure*}

Having established their structural, mechanical, and electronic properties, we next evaluate the photocatalytic water-splitting performance of the InXPbP systems. In Figs.~\ref{fig:band-edge}(a--c), we illustrate the band-edge alignments of the InSPbP, InSePbP, and InTePbP monolayers, respectively. The CBM and VBM positions are aligned with respect to the vacuum level and compared with the H$^+$/H$_2$ ($-4.44$~eV) and O$_2$/H$_2$O ($-5.67$~eV) redox potentials at $\text{pH} = 0$, allowing a direct evaluation of their suitability for overall water splitting. Owing to the intrinsic asymmetry of the Janus structure, the X (S, Se, Te) and P atomic layers occupy the $(\bar{1}00)$ and $(100)$ surfaces, respectively. It should be noted that the conventional criterion of $E_g > 1.23$~eV is not always strictly required for Janus photocatalysts~\cite{Thanh2023}. Owing to the intrinsic built-in electric field, photogenerated electrons and holes can migrate toward opposite surfaces, where the H$_2$ and O$_2$ evolution reactions occur separately. This unique feature enables efficient utilization of photoexcited carriers and allows Janus InSPbP, InSePbP, and InTePbP monolayers to satisfy the requirements for photocatalytic water splitting, even when the band gap approaches or falls below 1.23~eV.

To further assess the photocatalytic performance of the InXPbP systems, we calculate the solar-to-hydrogen (STH) efficiency, $\eta_\text{STH}$, using the following expression~\cite{thanh2023janus,bao2025rational}:
\begin{equation}
  \label{eq:sht}
\eta_\text{STH}= \eta_{abs}\times \eta_{cu}\times\frac{\int_{0}^{\infty}P(h\omega)d(h\omega))}{\int_{0}^{\infty}P(h\omega)d(h\omega)+\Delta V\int_{E_{g}}^{\infty}\frac{P(h\omega)}{h\omega}d(h\omega)},
\end{equation}
where $\eta_{abs}$ and $\eta_{cu}$ refer to the light absorption and carrier utilization efficiencies, respectively, and $E_g$, $\Delta V$, and $P(h\omega)$ denote the band gap, macroscopic potential difference, and AM1.5G solar flux at photon energy $h\omega$, respectively.

The light absorption $\eta_{abs}$ is calculated as the following equation:
\begin{equation}
  \label{eq:abs}
\eta_{abs}=\frac{\int_{E_{g}}^{\infty }P(h\omega)d(h\omega)}{\int_{0}^{\infty }P(h\omega)d(h\omega)}.
\end{equation}

The carrier utilization $\eta_{cu}$ is determined as: 
\begin{equation}
  \label{eq:cu}
\eta_{cu}=\frac{\Delta \Phi \int_{E_{min}}^{\infty }\frac{P(h\omega)}{h\omega}d(h\omega)}{\int_{E_{g}}^{\infty }P(h\omega)d(h\omega)},
\end{equation}
where $\Delta \Phi$ = 1.23 eV for water splitting and 
\begin{equation}
  \label{eq:emin}
E_{min} =
  \begin{cases}
    E_{g}       & \quad \text{for } \chi(\text{H}_{2}) \geq 0.2,\chi(O_{2}) \geq 0.6\\
    E_{g} + 0.2-\chi(\text{H}_{2})  & \quad \text{for } \chi(\text{H}_{2}) <  0.2,\chi(\text{O}_{2}) \geq 0.6 \\
    E_{g} + 0.6-\chi(\text{O}_{2})  & \quad \text{for } \chi(\text{H}_{2}) \geq 0.2,\chi(\text{O}_{2}) <  0.6 \\
    E_{g} + 0.8-\chi(\text{H}_{2})-\chi(\text{O}_{2}) & \quad \text{for } \chi(\text{H}_{2}) <  0.2,\chi(\text{O}_{2}) <  0.6
  \end{cases},
\end{equation}
where $\chi(\mathrm{H}_2)$ and $\chi(\mathrm{O}_2)$ represent the overpotentials for the hydrogen and oxygen evolution reactions, defined as the energy differences between the CBM and $\mathrm{H}^+/\mathrm{H}_2$ level, and between the VBM and $\mathrm{O}_2/\mathrm{H}_2\mathrm{O}$ level, respectively (see Fig.~\ref{fig:band-edge}).

\begin{table*}[t]
\centering
\caption{Overpotentials $\chi(\text{H}_{2})$ and $\chi(\text{O}_{2})$ (eV), light absorption efficiency $(\eta_{abs})$, carrier utilization $(\eta_{cu})$, and STH efficiency $(\eta_\text{STH})$}
%\vspace{0.5cm}  % Thêm khoảng cách 0.5cm
%\small
%\footnotesize  % cỡ chữ trong bảng
\renewcommand{\arraystretch}{1.5}
\begin{tabular}{c c c c c c c c}
\hline\hline
Materials & $\chi(\text{H}_{2})$ & $\chi(\text{O}_{2})$ &$\eta_{abs}(\%)$ & $\eta_{cu}(\%)$ & $\eta_\text{STH}(\%)$  \\ 
\hline    
InSPbP & 0.03 & 0.16 & 75.79 & 30.98 & 21.67\\  
InSePbP & 0.18 & 0.24 & 72.40 & 40.97 & 26.03 \\ 
InTePbP & 0.10 & 0.06 & 93.64 & 43.52 & 29.83\\  
\hline\hline
\end{tabular}
\label{table:STH}
\end{table*}    

The calculated values of $\chi(\mathrm{H}_2)$ and $\chi(\mathrm{O}_2)$ for the Janus InSPbP, InSePbP, and InTePbP monolayers are presented in Table~\ref{table:STH}. The calculated STH efficiencies of Janus InSPbP, InSePbP, and InTePbP monolayers are 21.67\%, 26.03\%, and 29.83\%, respectively, indicating that InTePbP exhibits the best photocatalytic performance among the three systems. This trend can be partly attributed to the increasing electrostatic potential difference ($\Delta V$), which enhances the intrinsic out-of-plane electric field and promotes the separation of photogenerated electron--hole pairs. To further understand this behavior, we analyze the corresponding overpotentials $\chi(\mathrm{H}_2)$ and $\chi(\mathrm{O}_2)$. All three materials exhibit positive values of $\chi(\mathrm{H}_2)$ and $\chi(\mathrm{O}_2)$, confirming that both hydrogen and oxygen evolution reactions are thermodynamically feasible. In particular, InSePbP shows the largest overpotentials, indicating a strong driving force for the redox reactions; however, excessively large overpotentials may lead to increased energy losses. In contrast, InTePbP exhibits relatively small positive overpotentials for both HER and OER, which is consistent with its high STH efficiency. These results suggest that the combined effects of the intrinsic electric field (associated with $\Delta V$) and favorable overpotentials contribute significantly to the enhanced photocatalytic performance of InTePbP. The calculated STH efficiency of the Janus InTePbP monolayer (29.83\%) exceeds the conventional theoretical limit of 18\%~\cite{fu2018intrinsic} and is higher than those reported for Pd$_4$Se$_3$Te$_3$ (23.8\%)~\cite{luo2021pd4s3se3}, WSSe (11.7\%)~\cite{ju2020janus}, GaInTe$_2$ (21.45\%)~\cite{guo2025janus}, and is comparable to that of Janus Ge$_2$SSe (28.78\%)~\cite{thanh2023janus}, and B$_2$P$_6$ (28\%)~\cite{sun2020b2p6}. For completeness, the light absorption efficiency, carrier utilization efficiency, and STH efficiency of Janus InXPbP monolayers are summarized in Table~\ref{table:STH}. Interestingly, the variation in STH efficiency follows the same order as $\alpha_R$, increasing from InSPbP to InSePbP and further to InTePbP. This correlation suggests that stronger Rashba spin splitting may contribute to improved carrier dynamics by suppressing electron--hole recombination. Furthermore, this behavior is consistent with the increasing $\Delta V$, which enhances the intrinsic out-of-plane electric field in Janus structures. The strengthened internal electric field not only promotes charge separation but also amplifies the Rashba effect, thereby synergistically improving the photocatalytic performance.

%In contrast, InTePbP exhibits small overpotentials, providing a balanced driving force for both reactions, which is beneficial for achieving higher STH efficiency~\cite{xxx}. 

The high STH efficiency of InTePbP can also be attributed to its excellent light absorption efficiency $\eta_{abs}$ (93.64\%), which originates from its relatively small band gap (0.76 eV). According to Eq.~\eqref{eq:abs}, a smaller $E_g$ leads to a higher $\eta_{abs}$. It should be noted that this expression is based on an ideal step-function approximation of the absorbance, in which photons with energies below $E_g$ are not absorbed, whereas those above $E_g$ are fully absorbed~\cite{tiedje1984limiting}. Consequently, $\eta_{abs}$ may be overestimated, particularly near the absorption edge. Nevertheless, the calculated values still provide useful insights, as the approximation is more reliable at higher photon energies.

\begin{figure*}[t]
  \centering \includegraphics[clip,width=14cm]{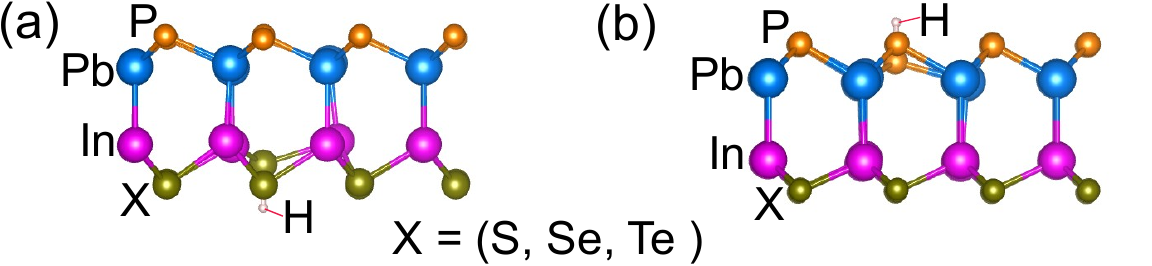}
  \caption{Side views of the optimized hydrogen adsorption configurations on Janus InXPbP: (a) X-terminated surface and (b) P-terminated surface.}
  \label{fig:model-shg}
\end{figure*}

To further assess the thermodynamic stability of the system, we evaluate the hydrogen adsorption free energy ($\Delta G_\mathrm{H}$), which is given by~\cite{thanh2023janus,bao2025rational,norskov2005trends}:
\begin{equation}
\label{eq:delta_G}
\Delta G_{\mathrm{H}} = \Delta E_{\mathrm{H}} + 0.24,
\end{equation}
where $\Delta E_H$ denotes the differential hydrogen adsorption energy, calculated as
\begin{equation}
  \label{eq:delta_H}
\Delta E_{\mathrm{H}} = E_{\mathrm{InXPbP+H}} - E_{\mathrm{InXPbP}} - \frac{1}{2}E_{\mathrm{H_{2}}},
\end{equation}
where $E_{\mathrm{InXPbP+H}}$ and $E_{\mathrm{InXPbP}}$ represent the total energies of the monolayer with and without an adsorbed H atom, respectively, and $E_{\mathrm{H_2}}$ corresponds to the energy of an isolated H$_2$ molecule. A $3 \times 3 \times 1$ supercell are employed to ensure convergence of the adsorption energy and minimize interactions between periodic images. The corresponding side view of the adsorption configurations of the X- and P-terminated surfaces is illustrated in Fig.~\ref{fig:model-shg}.

%Furthermore, based on the Nernst equation, the Gibbs free energy depends on pH as follows~\cite{chaoui2025novel}. This correction is essential for evaluating the pH-dependent hydrogen evolution reaction activity.

\begin{table}[t]
\centering
\caption{Gibbs free energy of hydrogen adsorption ($\Delta G_{\mathrm{H}}$, in eV) for different surfaces of Janus InXPbP monolayers.}
\begin{tabular}{lcccccc}
\hline\hline
\multirow{2}{*}{Material} & \multicolumn{2}{c}{InSPbP} & \multicolumn{2}{c}{InSePbP} & \multicolumn{2}{c}{InTePbP} \\
\cline{2-7}
& S-side & P-side & Se-side & P-side & Te-side & P-side \\
\hline
$\Delta G_H$ (eV) & 1.19 & 0.20 & 1.37 & 0.07 & 1.34 & $-0.18$ \\
\hline\hline
\end{tabular}
\label{tab:deltaG}
\end{table}

In Table~\ref{tab:deltaG}, we list $\Delta G_H$ on the X- and P-terminated surfaces of the InXPbP systems. The results reveal a pronounced surface dependence of hydrogen adsorption behavior. For the X-terminated surfaces (X = S, Se, and Te), the $\Delta G_H$ values are large and positive (1.19--1.37 eV), indicating energetically unfavorable hydrogen adsorption. Consequently, these surfaces are expected to be catalytically inactive toward the hydrogen evolution reaction (HER), as the weak interaction between hydrogen and the surface hinders the adsorption of reaction intermediates. The obtained $\Delta G_{\mathrm{H}}$ values are slightly lower than those reported for Janus $\gamma$-Ge$_2$SSe (1.54 eV)~\cite{thanh2023janus} and $\beta$-PtSSe (1.47 eV)~\cite{jamdagni2022photocatalytic}, and are comparable to that of AlS (1.24 eV)~\cite{haman2022janus}, further confirming the weak hydrogen adsorption strength on the X-terminated surfaces. In contrast, the P-terminated surfaces exhibit significantly smaller $\Delta G_{\mathrm{H}}$ values of 0.20, 0.07, and $-0.18$ eV for InSPbP, InSePbP, and InTePbP, respectively, indicating a much stronger affinity for hydrogen and enhanced catalytic activity. These results identify the P-terminated surfaces as the dominant active sites for HER. Among the three systems, InSePbP exhibits the most favorable adsorption strength, with $\Delta G_{\mathrm{H}} = 0.07$ eV, which is closest to the thermoneutral condition ($\Delta G_{\mathrm{H}} \approx 0$). Although InTePbP exhibits a negative $\Delta G_{\mathrm{H}}$ at pH = 0, indicating relatively strong hydrogen binding, the adsorption strength can be effectively tuned by adjusting the solution pH. According to the Nernst relation~\cite{norskov2004origin,chaoui2025novel}, the hydrogen adsorption free energy shifts by approximately 0.059 eV per pH unit, yielding a value of about 0.23 eV at $\text{pH} = 7$. This shift toward near-thermoneutral adsorption conditions is expected to facilitate both hydrogen adsorption and desorption, thereby enhancing the overall HER activity.

\subsection{Optical Properties}

\begin{figure*}[t] 
  \centering 
  \includegraphics[width=\linewidth]{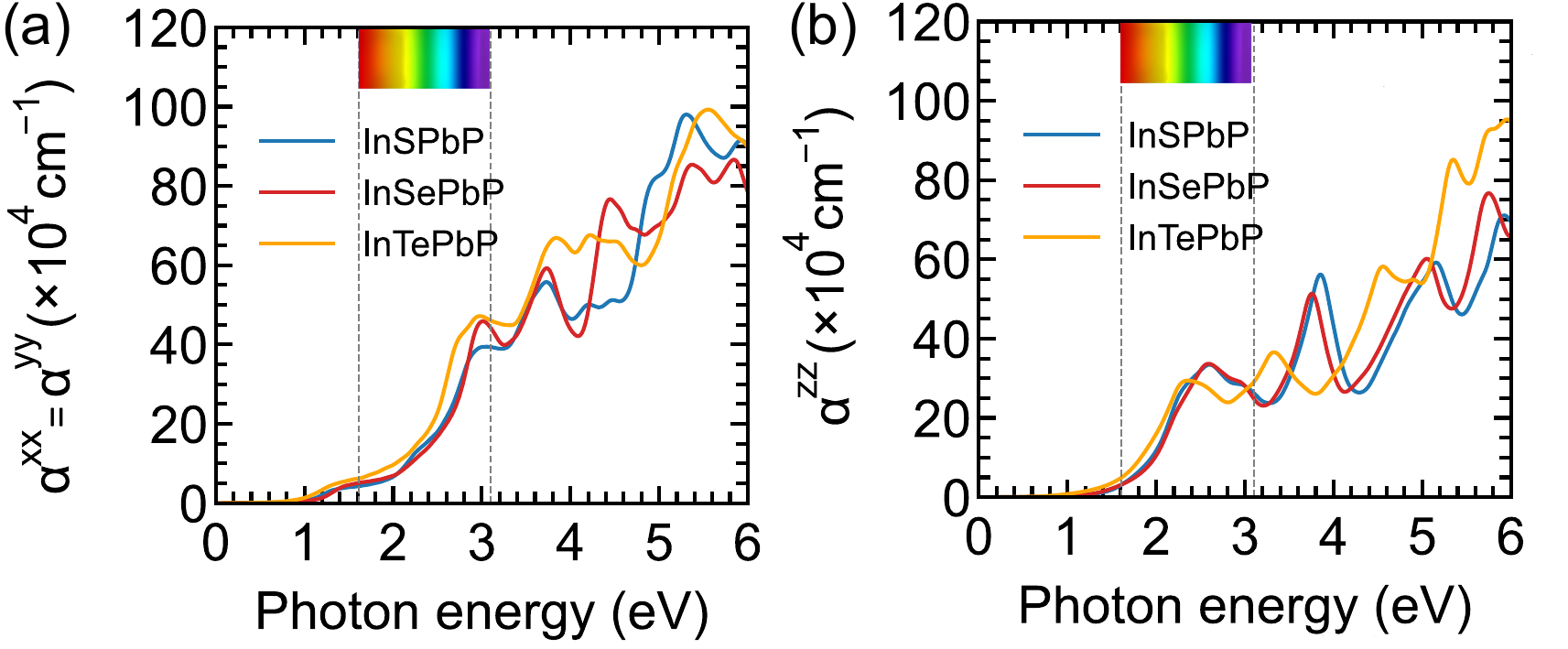}
\caption{\label{fig:optical} Optical absorption coefficients of Janus InXPbP monolayers along the (a) in-plane and (b) out-of-plane directions. The shaded region highlights the visible light range (1.61–3.10 eV).}
  \label{fig:optic}
\end{figure*}

In addition to catalytic activity, the optical properties are studied to further evaluate the photocatalytic performance of Janus InXPbP monolayers, in which the optical absorption coefficient $\alpha(\omega)$ is calculated by the complex dielectric function $\varepsilon(\omega)$ using the following relation~\cite{nguyen2022QE}:
\begin{equation}
\label{eq:absorption}
\alpha(\omega) = \frac{\sqrt{2},\omega}{c}
\left[ \sqrt{\varepsilon_{1}^{2}(\omega) + \varepsilon_{2}^{2}(\omega)} - \varepsilon_{1}(\omega) \right]^{1/2},
\end{equation}
where $\varepsilon_{1}(\omega)$ and $\varepsilon_{2}(\omega)$ denote the real and imaginary components of the dielectric function, respectively, and $c$ and $\omega$ represent the speed of light in vacuum and the angular frequency of the incident radiation. Figures~\ref{fig:optical}(a) and (b) show the optical absorption coefficient $\alpha(\omega)$ of Janus InXPbP monolayers obtained from GGA calculations with a scissor correction using HSE band gaps (see Table~\ref{tab:table3}) over the photon energy range of 0--6 eV. We note that, for these 2D systems, the dielectric functions are rescaled by a factor of $h/d_{0}$, where $h = 30$~\AA\ denotes the unit-cell thickness and $d_{0}$ corresponds to the effective monolayer thickness (see Table~\ref{tab:table1}).

We obtain that $\alpha^{xx}=\alpha^{yy}$ in the in-plane directions is significantly larger than its out-of-plane counterpart, $\alpha^{zz}$, throughout the visible-light region, indicating pronounced optical anisotropy in the layer 2D materials. Among the investigated systems, InTePbP exhibits the strongest in-plane optical response, with a maximum absorption coefficient of $47.1 \times 10^{4}$ cm$^{-1}$ at a photon energy of approximately 2.9 eV. This value exceeds those reported for monolayer $\gamma$-GeS ($37 \times 10^{4}$ cm$^{-1}$)~\cite{van2022effects}, SGa--PbP ($23 \times 10^{4}$ cm$^{-1}$)~\cite{chaoui2025novel}, and Janus TMD monolayers ($15 \times 10^{4}-30 \times 10^{4}$ cm$^{-1}$)~\cite{van2020first}, highlighting the outstanding light-harvesting capability of InTePbP. For the out-of-plane polarization, the maximum absorption coefficient reaches approximately $34.1 \times 10^{4}$ cm$^{-1}$, observed in both InSPbP and InSePbP at photon energies around 2.6 eV. Although smaller than the in-plane response, this substantial absorption coefficient indicates that optical transitions associated with perpendicular polarization remain highly active, enabling efficient light harvesting along the $z$ direction. This anisotropic optical response originates from the intrinsic layered nature of the Janus structures and the direction-dependent electronic transitions, which enhance in-plane light--matter interactions. Combined with their favorable STH efficiencies, these optical characteristics further underscore the potential of the InXPbP systems for optoelectronic and photocatalytic water-splitting applications.

\section{Conclusions}
In summary, we have systematically investigated the structural, mechanical, electronic, Rashba spin-splitting, photocatalytic, and optical properties of Janus InXPbP (X = S, Se, Te) monolayers using first-principles calculations. All three monolayers are confirmed to be dynamically and mechanically stable. Their ideal strengths decrease from InSPbP to InTePbP, reflecting the weakening interatomic bonding with increasing chalcogen atomic size. Electronic structure calculations show that, without SOC, InSPbP is a direct-gap semiconductor, whereas InSePbP and InTePbP exhibit indirect band gaps. The inclusion of SOC drives a direct-to-indirect band-gap transition in InSPbP, while the indirect-gap nature of InSePbP and InTePbP remains unchanged. Moreover, all three systems exhibit Rashba spin splitting, with InTePbP displaying remarkably large Rashba parameters of 0.90 and 0.87~eV\AA\ near the VBM and CBM, respectively. The favorable band-edge alignments, strong visible-light absorption, and high STH efficiencies of 21.67\%, 26.03\%, and 29.83\% for InSPbP, InSePbP, and InTePbP, respectively, demonstrate the excellent photocatalytic performance of these monolayers. These findings highlight Janus InXPbP monolayers as promising multifunctional materials for spintronic, optoelectronic, and photocatalytic water-splitting applications.

%%%END OF MAIN TEXT%%%
%\section*{Supporting Information}
%#The Supporting Information is available free of charge at \url{https://pubs.acs.org/doi/xxx.}
%Figure of the total energy and temperature as a function of time at $T$ = 900 K, figure of the band structure of MoTe$_{2}$/PtS$_{2}$ heterobilayer ($\sqrt{3} \times 1$ supercell), figure of the total energy-strain relationship, figure of the calculated band energies, figure of the temperature-dependent relaxation time at several strain values, figure of the Seebeck coefficient $S$, electrical conductivity $\sigma$, electronic thermal conductivity $\kappa_{el}$, and power factor PF are plotted as function of the chemical potential $\mu$ at $T$= 500 K, 700 K and 900 K.

\section*{Acknowledgements}
V.V.T. and N.M.Q. acknowledge the Vietnam’s National Foundation for Science and Technology Development (NAFOSTED) with No. 107.02-2025.19; N.T.H. acknowledges funding from National Taiwan University under Grant No. NTU-NFG-114L74135 and the Catalyst for Change Program of the University Academic Alliance in Taiwan (UAAT), Ministry of Education, Taiwan (UAAT-CFC-11401).

%\section*{References}
%\bibliographystyle{elsarticle-num}
%\bibliography{thanh-ref}
\bibliography{References}
%\end{mcitethebibliography}

\end{document}